\documentclass[12pt, preprint]{aastex}

\def\deg{\ifmmode^\circ\else$^\circ$\fi}

 \def\mic{~$\mu$m}

\def\arcs{\ifmmode {''}\else $''$\fi}
\def\arcm{\ifmmode {'}\else $'$\fi}
\def\kms{km s$^{-1}$}
\def\parcs{\sa=.07em \sb=.03em
     \ifmmode $\rlap{.}$^{\scriptscriptstyle\prime\kern -\sb\prime}$\kern -\sa$
     \else \rlap{.}$^{\scriptscriptstyle\prime\kern -\sb\prime}$\kern -\sa\fi}
\def\parcm{\sa=.08em \sb=.03em
     \ifmmode $\rlap{.}\kern\sa$^{\scriptscriptstyle\prime}$\kern-\sb$
     \else \rlap{.}\kern\sa$^{\scriptscriptstyle\prime}$\kern-\sb\fi}

\def\Msun{\mbox{M$_{\odot}$}}
\def\Myr{\Msun/yr}
\def\kp{{\rm K}$^{\prime}$}

\def\han {\mbox{{\rm H}$\alpha$}}
\def\ha{\han}
\def\hb{\mbox{{\rm H}$\beta$}}

\def\h0{\mbox{H$_0$}}
\def\ergcm2s{\ifmmode {\rm\,ergs\,cm^{-2}\,s^{-1}}\else
                ${\rm\,ergs\,cm^{-2}\,s^{-1}}$\fi}

\def\spose#1{\hbox to 0pt{#1\hss}}
\def\simlt{\mathrel{\spose{\lower 3pt\hbox{$\mathchar"218$}}
     \raise 2.0pt\hbox{$\mathchar"13C$}}}
\def\simgt{\mathrel{\spose{\lower 3pt\hbox{$\mathchar"218$}}
     \raise 2.0pt\hbox{$\mathchar"13E$}}}
\def\lsim{\rlap{$<$}{\lower 1.0ex\hbox{$\sim$}}}
\def\gsim{\rlap{$>$}{\lower 1.0ex\hbox{$\sim$}}}

\begin{document}

\title{The Counter Arc to MS1512-cB58 and a Companion Galaxy\footnotemark}

\footnotetext[1]{
Data presented herein were obtained at the W.M. Keck Observatory, 
which is operated
as a scientific partnership among the California Institute of Technology, 
the University of California and the National Aeronautics and Space 
Administration.  The Observatory was made possible by the
generous financial support of the W.M. Keck Foundation.
}

\author{
Harry I. Teplitz,\altaffilmark{2} 
Matthew A. Malkan,\altaffilmark{3} \& Ian S. McLean\altaffilmark{3}
}
\altaffiltext{2}{Spitzer Science Center, MS 220-6,
California Institute of Technology, Pasadena, CA 91125. Electronic mail:
hit@ipac.caltech.edu}

\altaffiltext{3}{Department of Physics \& Astronomy, UCLA, Los Angeles, CA 90095.}

\begin{abstract}
  
  We present near-infrared spectra of ``A2'', the primary counter arc
  to the gravitationally lensed galaxy MS1512-cB58. The spectra show
  redshifted \ha, [NII], [OIII], and \hb\ at $z=2.728\pm 0.001$.  We
  observe the same \ha/[OIII] ratio as cB58, which together with the
  redshift confirms that A2 is indeed another image of a single
  background galaxy. Published lensing reconstruction reports that A2
  is a magnification of the entire source, while cB58 is an image of
  only a part. At marginal significance, A2 shows higher line to
  continuum ratios than cB58 (by a factor of $\sim 2$), suggesting a
  non-uniform ratio of young to old stars across the galaxy. We
  observe a second emission line source in the slit.  This object,
  ``W5'', is predicted to be a lensed image of another galaxy at a
  redshift similar to cB58.  W5 is blueshifted from cB58 by $\sim
  400$\ \kms\ and has a significantly lower \ha/[OIII] ratio,
  confirming that it is an image of a different background galaxy in a
  group with cB58.  The \ha\ emission line in W5 implies a star
  formation rate of 6 $\mbox{M}_{\odot}\ \mbox{yr}^{-1}$\ ($H_0 = 70$
  km s$^{-1}$\ Mpc$^{-1}$, $\Omega_M=0.3, \Omega_{\Lambda}=0.7$),
  after correcting for lensing magnification.

\end{abstract}


\section{Introduction}

Most of the detailed spectral information on $z>2$~ galaxies has come
from observation of the rest-frame ultraviolet redshifted into the
optical passband.  The Lyman Break Galaxies (LBGs; Steidel et al.
1996, 2003) are strongly starbursting galaxies, and in principle may
be the tracers of the global star formation history of the universe
(Madau et al. 1998) if the effects of dust extinction on the UV
continuum can be quantified.  The average attenuation from dust has
been suggested to be between a factor of $<2$\ and 10 (Pettini
et al.\ 1998; Trager et al. 1997).  Even with the advent of IR
spectrographs on large telescopes, only a few dozen $z>2$\ starbursts
have been spectroscopically observed in the rest frame optical (e.g.
Teplitz et al. 2000a, 2000b; Kobulnicky \& Koo 2000; Pettini et al.\ 
1998, 2001; Lemoine-Busserolle et al.\ 2003; Erb et al.\ 2003).  

MS1512-cB58 (hereafter cB58) is a strongly lensed, but otherwise
typical, star-forming galaxy at z=2.729 (Yee et al.\ 1996; Seitz et
al.\ 1998, hereafter S98).  Its magnification by a factor of $\sim
30-50$\ makes it the apparently brightest LBG known.  Spectra have
been obtained in the rest-frame UV (Pettini et al.\ 2000) and optical
(Teplitz et al.\ 2000b, hereafter T00).  Lensing reconstruction
identifies another faint source in the same field as a counter arc to cB58
(S98). This object, designated ``A2'' by S98, is a magnification of a different part of the
same background source.  Thus the gravitational lensing provides a
means to spatially resolve the observations of this intrinsically faint
source.  Reconstruction shows that cB58 is a magnification of only
half of the source galaxy, including the nucleus, and that A2 is a
magnification of  the entire source.

The lensing model also identifies several other faint objects as
probable arclets that are magnified by the foreground cluster (see S98 and their Figure 1).  At least
four background objects are likely to be lensed: the A source,
magnified as cB58 and A2; the B source, possibly at $z\sim 3$; the C
source, most likely at a redshift slightly larger than cB58; and the W
source, which should have a redshift similar to cB58.  The W system of
 consists of five arclets:  three sources with ``shrimp-like'' morphologies in
the Hubble Space Telescope (HST) imaging; a fourth, very faint source; and a 
probable 5th image with an uncertain predicted location.

In this paper, we present near-infrared (NIR) spectra and \kp\ imaging
of A2 and the probable fifth image of the W source, which we call W5.
The data were obtained with the NIRSPEC instrument on the Keck II
telescope.  We also compare our new observations with archival HST imaging
of the objects.  Throughout the paper, we will adopt a flat,
$\Lambda$-dominated universe ($H_0 = 70$ km s$^{-1}$\ Mpc$^{-1}$,
$\Omega_M=0.3, \Omega_{\Lambda}=0.7$).

\section{Observations and Data Reduction}

\subsection{Spectroscopy}

On 4 April 2002, we obtained the NIR spectrum of A2, using NIRSPEC
(McLean et al.\ 1998, 2000) on the 10-m W. M. Keck II telescope.
Observations were taken in the low resolution, long slit mode, using a
four pixel wide slit (0.76\arcs).  Spectra were obtained in both the H
and K bands (1.5-2.5 $\mu$m).  In this mode, the instrumental
resolution is $R\sim 1000$.  The weather was photometric and seeing
was very good during the observations, with point source full width at
half maximum (FWHM) of 0.4\arcs\ or better throughout.  The spectra
were obtained with the N6 and N7 filters, corresponding roughly to the H- and K
bands respectively.  Individual integration times were 600 seconds.  The
telescope was nodded along the slit between exposures to facilitate sky
subtraction.  The total integration times were 1800 and 3600 seconds
in the H- and K-bands respectively.

\begin{figure}
\plotone{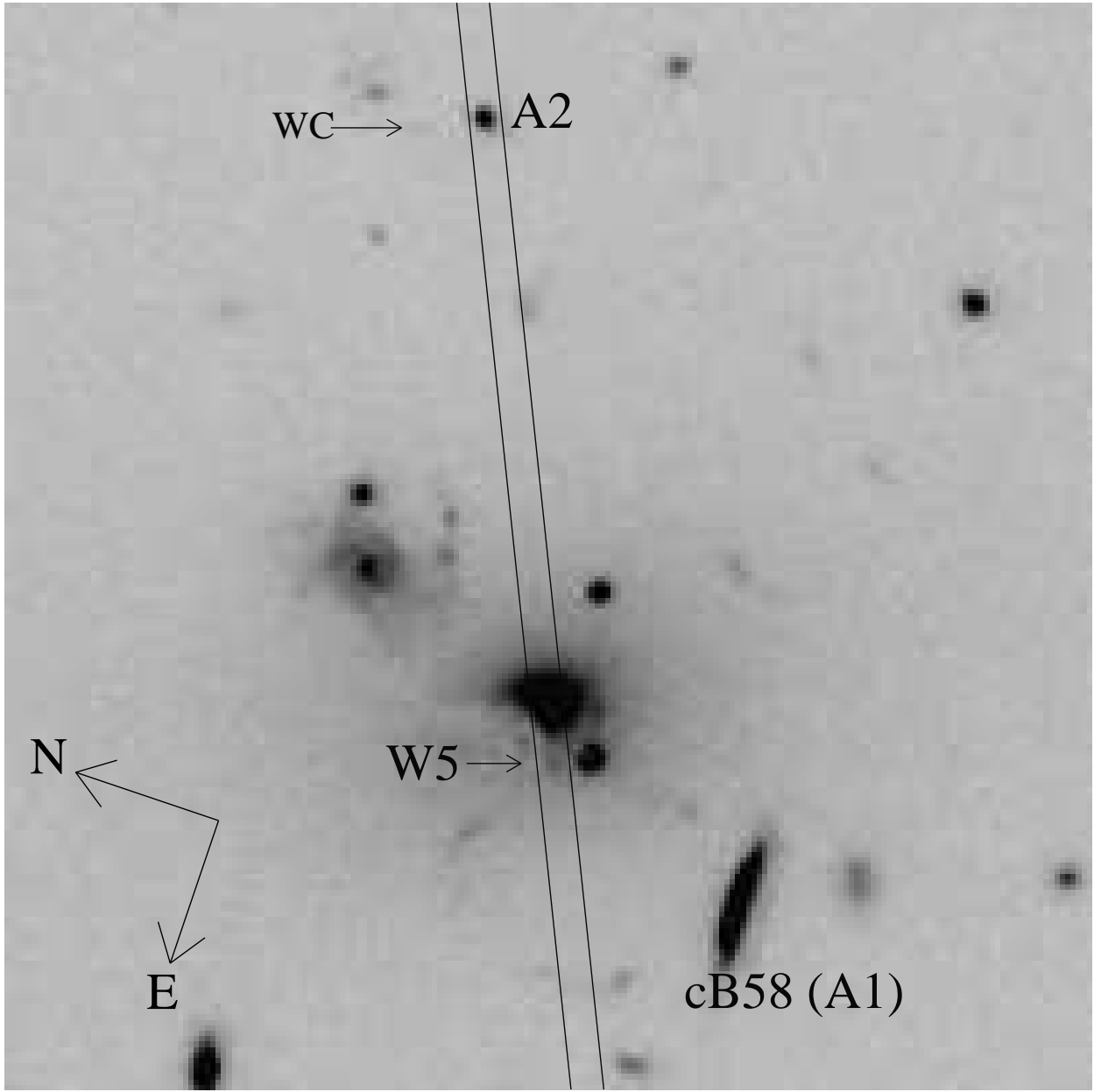}
\caption{\label{fig: fc} Orientation of the NIRSPEC slit.  The slit is plotted over a $20\arcs \times
20\arcs$\ section of the F675W WFPC2 image.}
\end{figure}

The slit was rotated to a position angle chosen to observe both a bright reference
object and A2.  The reference object facilitates registration of
nodded observations and allows us to monitor the guiding during the
observation.  By choosing the cD galaxy at the center of the MS1512
cluster for our reference, we also placed W5 in the slit.  The
position angle of the slit was 294.58\deg\ East of North.  Figure
\ref{fig: fc}\ shows the orientation and location of the slit.

To reduce systematic uncertainties, we observed the same calibration
stars used by T00 for atmospheric extinction correction and flux
calibration. These stars are a G star for the H-band (PPMM 130690) and
an A2V star for the K-band (HR5569).  Halogen flat fields and spectra
of argon and neon arc lamps were taken immediately after the
target observations and before moving the mechanisms, in order that the
calibration most closely reproduce conditions during the
observations.

Data were reduced following the procedures in T00 and Teplitz et al.\ 
(2000a).  Standard IDL and IRAF\footnote[4]{IRAF is distributed by
  NOAO, which is operated by AURA Inc. under contract to the NSF.}
routines were used for flat fielding and fixing bad pixels.
Rectification of the 2D spectra was performed using custom IDL
software, which uses linear interpolation to independently rectify the
spatial and spectral dimensions.  The output pixel scale is
4.16\AA/pix, and unresolved arc lines have $\mbox{FWHM}\sim 15$\ \AA,
for a final resolution of $R\sim 1050$\ at 1.6\mic.

For spectroscopy of faint objects in the near-IR, sky subtraction is a crucial
step.  The sky frame for each observation was made by scaling the other half of
the nodded pair.  The OH sky lines vary independently from each other and therefore
must be scaled individually.  Each nodded pair was rectified in two dimensions
and a scaling factor determined for each row (the spatial dimension).  Then the
image of scaling factors was de-rectified onto the original pixel scale and
applied to the unrectified sky image.  By subtracting the sky in original pixels,
we obtain better signal to noise, because the interpolation in the process is of the
scaling factor.  

The sky-subtracted spectra were then rectified in two dimensions and the nodded 
pairs registered.  The final 2D spectrum was a weighted mean of the individual
frames.   Weights were determined using variation in the brightness of the reference
object, under the assumption that such variation was the result of small guiding errors.
A 1D spectrum was extracted using an optimal extraction vector determined from the
calibration star continuum shape.  Extracted spectra were corrected for atmospheric
absorption using the standard stars.  Flux calibration also relied on the calculation
of flux density per data number per pixel in the standard star.

\subsection{Imaging}

NIRSPEC's slit-viewing camera, SCam, images a $46\arcs
\times 46\arcs$\ area around the entrance slit in parallel with
spectroscopic observations.  The camera has a $256\times 256$\ pixel
(PICNIC) HgCdTe array detector, with a plate scale of 0.18\arcs\ per
pixel.  During setup for the first spectrum, we used the SCam to
observe the MS1512 field in the \kp\ filter.  Four 60 second, dithered
exposures were taken, each consisting of 4 coadds of 15 seconds.  The
images were reduced using standard IRAF procedures.  A flat field was
created from the median of all \kp\ observations throughout the night.
We were unable to observe photometric standard
stars in the same filter during that night; however, we use the
image to measure differential photometry from published values for cB58 (Ellingson
et al.\ 1996).

WFPC2 optical images of the field were obtained from the Hubble Space
Telescope archive.  The data were processed by the Space Telescope
Science Institute as part of the new WFPC2 associations
archive\footnote[5]{http://archive.stsci.edu/hst/wfpc2/}, which
provides mosaiced data combined using the drizzle technique (Fruchter
\& Hook 2002).  Data were originally obtained by the programs GO 6003
(Saglia et al., see S98) and GO 6832 (Yee et al.).  Fully-reduced
mosaics were available in the archive in the F555W, F675W, and F814W
filters, corresponding roughly to the V, R, and I passbands, with
exposure times 19,800 seconds, 10,400 seconds, and 7600 seconds
respectively.

In both the WFPC2 and \kp\ images, the flux in cB58 is measured using
ellipse fitting to establish the curve of growth.  The flux is then
scaled to the photometry from Ellingson et al.\ (1996) to establish
relative zero points for the images. A2 is well separated from nearby
objects and we measure its flux using aperture photometry.  The W5
photometry is contaminated by its juxtaposition with the foreground cD galaxy.
We measure the flux from W5 in a small aperture (0.3 arcsecond
radius).  To estimate contamination from the cD galaxy,
we measure the same aperture in other regions of the cD galaxy at
equal distance from its center.  We then perform an aperture
correction to estimate W5's total flux, basing the correction on the
curve of growth of A2.  To check this procedure, we also measured W5 in
an image with a model of the cD galaxy subtracted (from S98) and
obtained results consistent at the 20\% level.

\subsection{Equivalent Widths}

In order to compare the new observations directly with the spectrum of
cB58, we have re-reduced that data to use the same flux calibration
technique.  In T00, the flux calibration relied on measurement of the
equivalent width using the continuum in the spectrum.  They performed a linear fit  
to the continuum across the entire wavelength range (either J, H or K) and
used the derived flux at the wavelength of each emission line in the calculation
of the equivalent width (EW).  This procedure was not fully explained in the text of T00.  

In the present data, EWs cannot be measured directly from the spectra,
because detection of the continuum is poor.  Instead, the EW of lines
in A2 is estimated as the ratio of the measured flux to the continuum
flux density in the photometry.  The flux density is normalized to the median 
wavelength of the filter, and the slope of the continuum is assumed to match that
measured in cB58.  H-band photometry is not yet
available for A2 or W5.  For A2, we assume that the H-K color is the
same as in cB58. The optical colors are of the two counterparts are
nearly identical.  This assumption introduces an unknown uncertainty into
our ability to compare the EW([OIII]) of A2 to cB58.

The K-band photometry of W5 has a low signal to noise ratio, and the
H-K color is not known, so EWs are not estimated for W5.  The lines
must have fairly high equivalent width to be detectable, given how
faint W5 is in the continuum.  We do not detect the continuum in the
spectra, but the lines are convincingly real.  Each line is detected
in each of the individual exposures of each nodded pair of spectra.

\subsection{Estimating the Uncertainties}

Each step in the calculation of the EW has an associated uncertainty:

First, we measure the line intensity in DN and convert to flux using
the spectrophotometric calibration from the standard star.  The
spectrophotometric calibration is uncertain due to our knowledge of
the standard star and the variation in the atmosphere between the
observations of the source and of the standard.  Together, we assume
those factors introduce a 10\% uncertainty.  The measurement of the
line is uncertain due to shot noise in the background and Poisson
noise in the source.  We measure these from the square root of an extracted
sky spectrum added to the object spectrum.  

Next, we divide by the flux density of the continuum under the line,
assuming the slope of the continuum in A2 matches that in cB58.  The linear
fit is uncertain at the 5\% level (based on repeated measurements of the
cB58 spectrum).  The linear fit is normalized by the photometric measurement,
which has less than 10\% uncertainty for cB58, but has 16\% uncertainty 
in A2.  

Finally, as noted above, the [OIII] measurement has additional (unquantified)
uncertainty due to the current lack of H-band photometry.

\section{Results}

\begin{figure}
\plotone{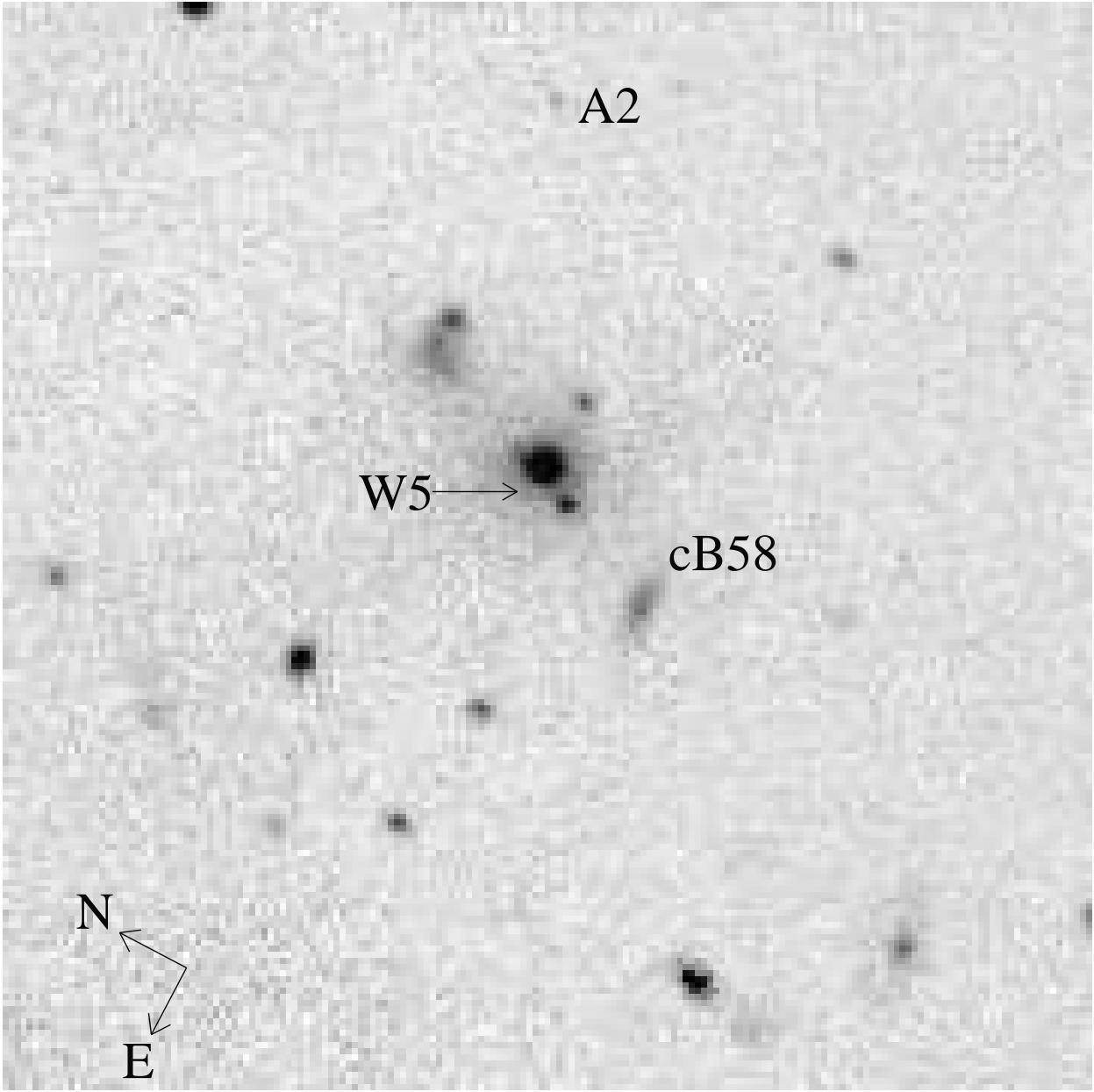}
\caption{\label{fig: kimage} The K-band image of the field, taken with the NIRSPEC Slit-viewing Camera.
The image is $32\arcs \times 32\arcs$.  The image is a at a position angle 9 degrees clockwise from Figure 1.}
\end{figure}

\begin{figure}
\plottwo{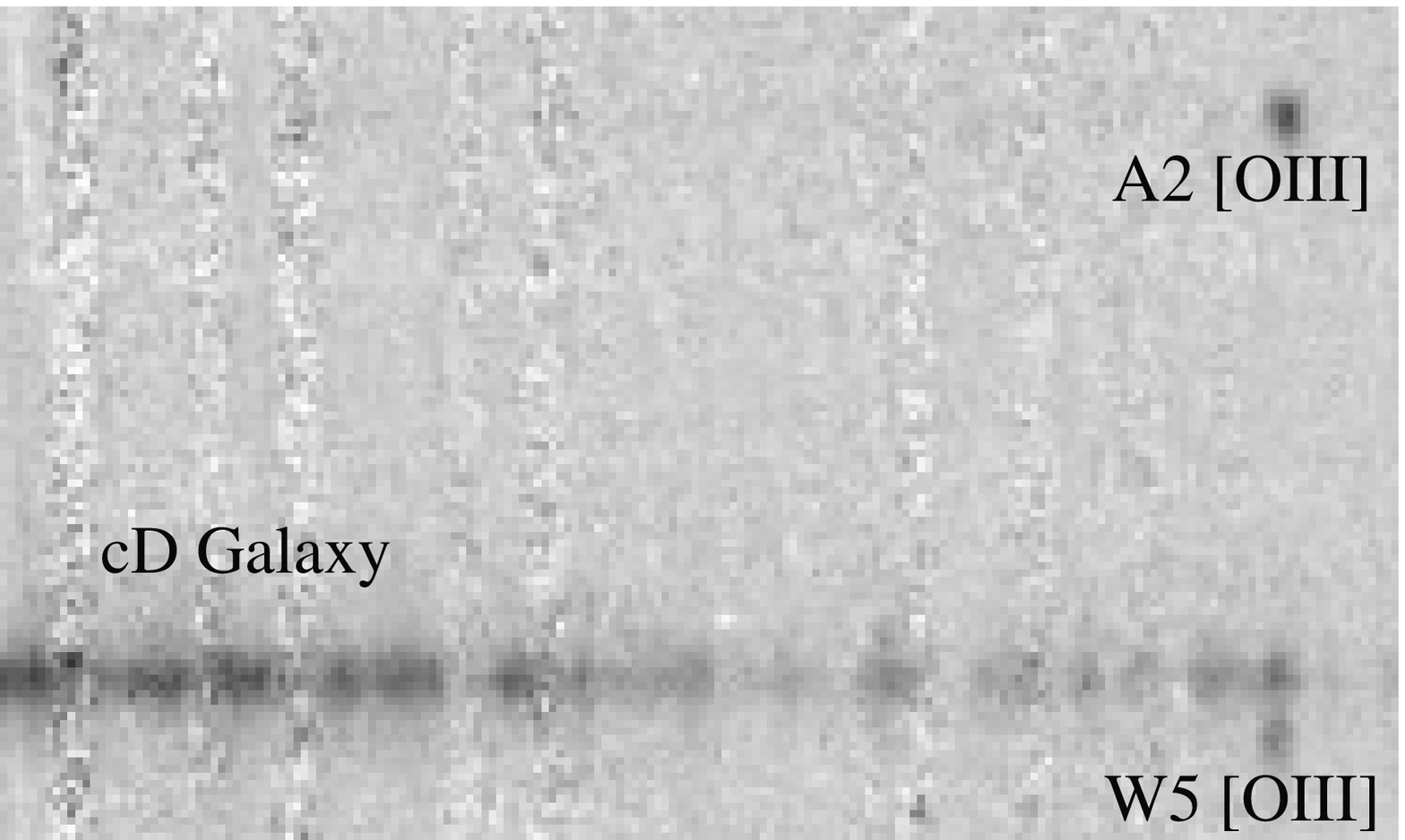}{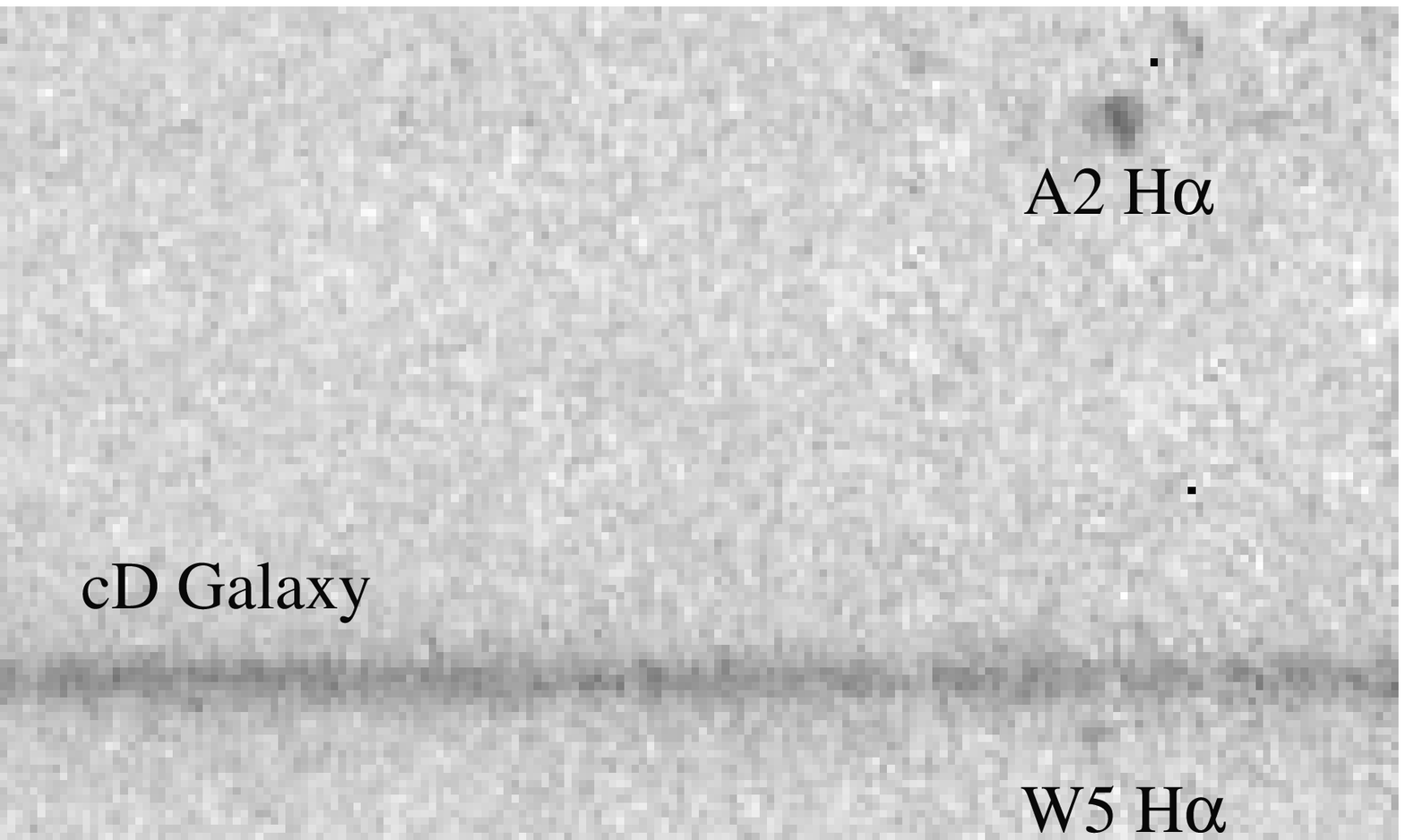}
\caption{\label{fig: 2d spectra} The two dimensional spectra in the H (left) and K (right) bands.  The
images show only a subset of the entire spectra, focusing on the region with strong emission lines.
The images show spectra regions covering 520\AA\ and 700\AA\ respectively.  In each image, the strong
W5 emission line is seen below the cD galaxy spectrum.  The variation in the cD spectrum is the result
of varying atmospheric transmission rather than spectral features.}
\end{figure}

\begin{figure}
\plotone{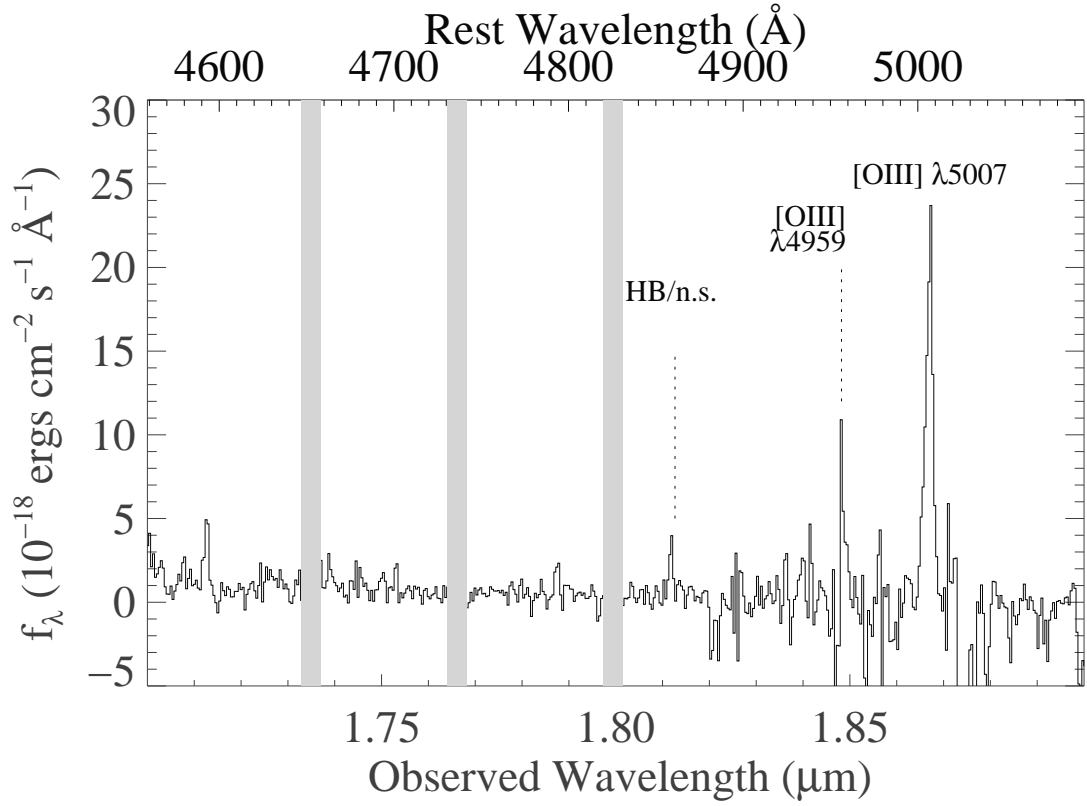}
\caption{\label{fig: 1d hspec} H-band spectrum of A2.  Detected emission lines are indicated.  Night sky lines with
particularly poor subtraction are shown by shaded regions.}
\end{figure}

\begin{figure}
\plotone{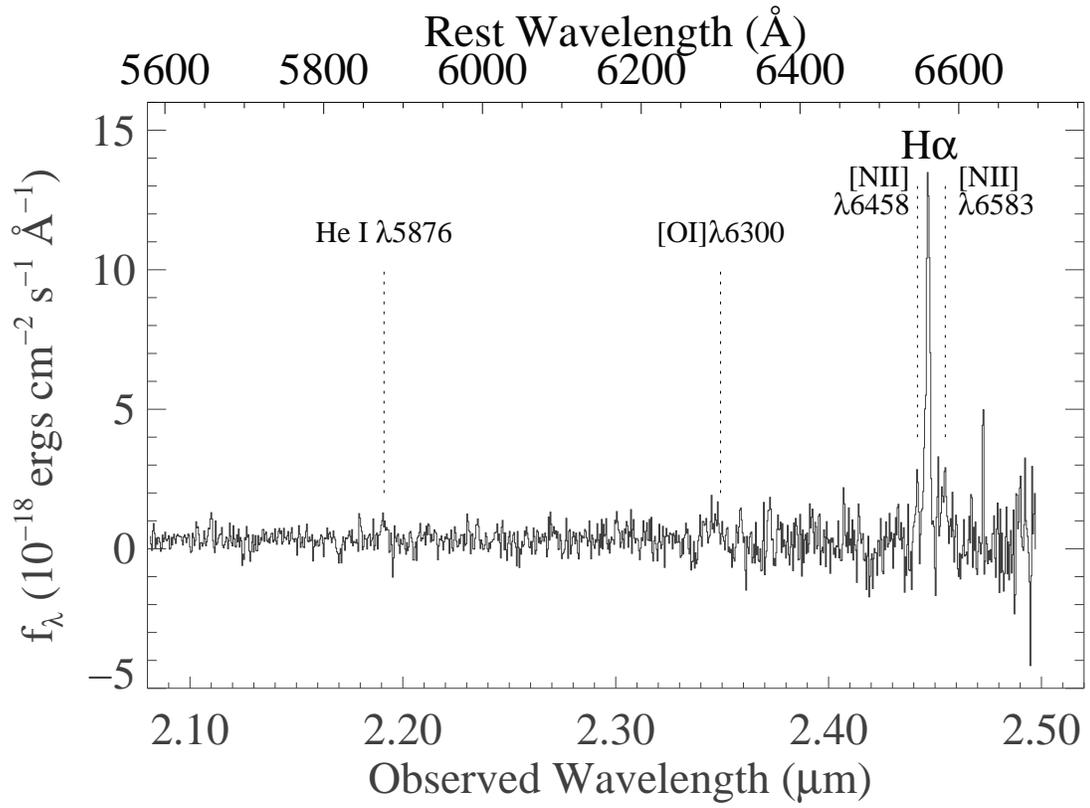}
\caption{\label{fig: 1d kspec} K-band spectrum of A2.  Detected emission lines are indicated.}
\end{figure}

Figure \ref{fig: kimage}\ shows the NIRSPEC-SCam K-band image of the MS1512
field.  Table \ref{tab: phot}\ lists the photometry measured in the
WFPC2 and NIRSPEC imaging.  Figure \ref{fig: 2d spectra}\ shows the
two dimensional NIRSPEC spectra; Figures \ref{fig: 1d hspec}\ and
\ref{fig: 1d kspec}\ show the extracted H and K spectra of A2 and W5.
We clearly detect the [OIII] doublet and H$\alpha$\ in both objects.
 H$\beta$\ falls partially on a night sky line.  Table
\ref{tab: emlines}\ gives the measured flux for the \ha\ and
[OIII] 5007\AA\ emission lines.

\subsection{Comparison of A2 and cB58}

The spectra confirm that A2 is a magnification of the same source
galaxy as cB58.  Their redshifts are nearly identical, $2.7288\pm
0.0007$\ compared to $2.7290\pm 0.0007$\ (T00).  The $V_{555}$-band WFPC2
photometry is consistent with the S98 measurements from separate
reduction of the same data.  The additional $I_{814}$- and K-band data confirm
the brightness and therefore that the magnification of A2 is $\sim
11$\ times less than that of cB58.

Any possible differential velocity between A2 and cB58 must be less
than 125 km s$^{-1}$.  This small velocity difference provides no
evidence of a systematic rotational velocity across the face of the
source galaxy.  Erb et al.\ (2003) observe rotational velocities of
$50<v_c<240$\ km s$^{-1}$\ from \ha\ in LBGs at $z\sim 2.3$.  Only the
largest values of $v_c$\ in their sample would be detectable with the
resolution we obtain in comparing A2 and cB58.  That is, if $v_c$\ 
were calculated only twice for each galaxy in the Erb et al.\ sample
(once averaged over the entire source, and once averaged of half the
source), then only the largest $v_c$\ would be detected.
Lemoine-Busserolle et al.\ (2003) observe velocity gradients of $\sim
240$\ km s$^{-1}$\ in spatially resolved NIR spectra of $z\sim 1.9$\ 
galaxies.  Gradients of that magnitude are detectable with our
resolution.

The limit on  rotational velocity in the A source is consistent
with the value inferred from the emission line widths.  The FWHM of
lines detected in cB58 suggests $\Delta v\sim 190$\ \kms, after
correcting for instrumental resolution.  None of the emission lines in
A2 are resolved in velocity.  Given the four-pixel entrance slit, this
velocity resolution limit is $\sim 250$\ km s$^{-1}$, which is consistent with
cB58.  The resolution of the A2 spectra is lower than in the cB58
observation, because we used a wider entrance slit for A2.

Comparison of the characteristics of A2 and cB58 provide spatially
resolved information about the source galaxy.  The ratio of the
strongest emission lines (\ha, [OIII]) is the same, within the
uncertainties, for the two arcs.  Better measurements are needed to
estimate metallicity from the A2 spectra.  Detections of \hb\ and
[OII] would allow us to infer the oxygen abundance from the ([OII] +
[OIII])/\hb\ ratio, which is often denoted $R_{23}$\ (Pagel et al.\ 
1979).  Nonetheless, the \ha:[OIII] ratio suggests that the
metallicity is the same in A2 as cB58, indicating no metallicity
gradient in the source galaxy.  A strong spatial variation of the
extinction could also cause a difference in the emission-line ratio,
so we conclude that the extinction has no strong gradient.

[NII] is marginally detected.  Its EW is consistent with cB58 as
measured by T00, but we cannot rule out a factor of two overestimate in
the previous measurement as suggested by Pettini et al. (2002).  The
nitrogen abundance sets a lower limit on the age of the galaxy if
it is high enough to require a secondary phase of metal production
(Kobulnicky \& Zaritsky 1999).  The Pettini et al.\ estimate suggests
that such a phase is not needed.

A possible difference between cB58 and A2 is that the EW of the strong
emission lines in A2 is larger than that observed in cB58.  This
difference is significant only at the 2.5$\sigma$\ level in the
current data, but is nonetheless intriguing.  We assume that the
continuum comes mostly from older stars and that the strong emission
lines are reprocessed radiation from young stars.  The difference in
EW, then, may result from a non-uniform distribution of recent star
formation across the face of the source galaxy.  Adopting this
explanation, we infer the relative the star formation rate (SFR) in
the two images of the A source.  From Kennicutt (1998): 
$SFR (\mbox{M}_{\odot}\ \mbox{yr}^{-1}) = 7.9\times 10^{-42} L(\ha)(\mbox{ergs s}^{-1})$.  
The SFR of A2 is 144 \Myr, only a factor
of four lower than cB58, compared to a magnification ratio of 11.  The
lensing reconstruction (S98) shows that cB58 is a magnification of
only half the galaxy, including most of the nuclear region.  It is
unlikely, therefore, that the difference is simply the result of
nucleus-centered star formation.

The apparent EW difference between A2 and cB58 is strikingly large.
Both images encompass at least half of the source galaxy. Thus, an EW
ratio of two would imply that the part of the galaxy not magnified
into cB58 has a line to continuum ratio that is three times larger of
the other half.  This difference would be surprising for a galaxy that
does not have an highly irregular morphology.  Although, there are
examples of intrinsically asymmetric starbursts at similar redshift
(Lowenthal et al.\ 1996), perhaps the results of mergers (Conselice et
al.\ 2003), there is no indication of such asymmetry in the HST image
of cB58.  In fact, S98 suggest that, contrary to our results, cB58
should be a ``zoom'' into the region of highest star formation
density.

\subsection{Confirmation of W5}

The location of W5 is consistent with the position of a lensing arc of
the W source galaxy as predicted by S98.  We convincingly measure \ha\ 
and [OIII], although the continuum is too faint for detection.  From
the optical emission lines, we infer a systematic redshift of
$2.724\pm 0.001$.  This redshift confirms the S98 prediction that the
W source is a different galaxy than the A source, although both are at
similar redshift.

The faintest component of the W system, WC (see Figure \ref{fig: fc}),
has a magnitude of $V_{555}=27.04\pm 0.6$\ (S98).  WC is too faint to
be seen in the NIRSPEC K-band image.  Extinction by the cD galaxy may
suppress the observed $V_{555}$-band flux of W5, but it is about a
magnitude brighter than WC, so it appears to be magnified by at least
that much.  The $R_{675}$-band detection of W5 is consistent with this
estimate, but the $R_{675}$-band fluxes of both W5 and WC are more
uncertain.  The one magnitude of magnification is small compared to
the factor of 10 by which the W1 component of the system is magnified
(S98).

From the magnification-corrected \ha\ emission line, we can infer SFR
of the W source.  A magnification factor of 2.5 suggests an unlensed
SFR of 6 $\mbox{M}_{\odot}\ \mbox{yr}^{-1}$.  It is impractical to
infer the SFR from the UV continuum of W5, given the extinction by the
cD galaxy.  We can estimate the redshifted 1500\AA\ continuum of the W
source from the F555W flux of its other lensed images.  The UV
continuum implies SFR$\sim 1.3~ \mbox{M}_{\odot}\ \mbox{yr}^{-1}$\ 
(Kennicutt 1998).  This factor of four difference between the SFRs
inferred from \ha\ and the UV, is the same as the difference observed
for cB58.

The \ha/[OIII] ratio in W5, $0.41\pm 0.13$, is significantly lower
than in the A source.  This ratio is an upper limit, as one might
expect there to be some absorption by the foreground cD galaxy.  It is
tempting to explain the difference in the \ha/[OIII] ratio between the
W5 and cB58 (i.e. the A source) as the result of lower intrinsic
reddening in the W source.  However, we have seen that the ratio of
SFRs inferred for the W source from UV-continuum and \ha\ is
consistent with the extinction in the cB58, E(B-V)$\sim 0.25$\ (T00;
Pettini et al.\ 2000; Baker et al.\ 2001).  Apart from reddening, the
stronger [OIII] emission in W5 could also be a sign that it has lower
metallicity than cB58.  Oxygen line strengths in sub-solar metallicity
galaxies can decrease with increasing oxygen abundances due to
cooling.  A lower limit on this effect occurs at $Z\sim 0.3Z_{\odot}$,
because the lack of oxygen atoms reduces the line strength (i.e.\ 
Kobulnicky et al.\ 1999).  The W source is likely to have relatively
low metallicity; most LBGs have oxygen abundances of several tenths of
solar (Pettini et al.\ 2001), and metallicity is typically lower in
less massive star forming galaxies (i.e. Melbourne \& Salzer 2002).
Finally, the [OIII] strength in W5 could be an indication of an active
nucleus.  However, we do not detect other lines that might be expected from a
narrow-line Seyfert (such as strong [OI] 6300\AA) or the presence of
broadened wings to the Balmer line.

We do not know the transverse separation of the A and W sources in the
spatial dimension, but we expect that it is smaller than the apparent
separation between the counter arcs.  The separation of cB58 and A2 is
15.3\arcs, and the widest separation in the W system is comparable.
At $z=2.72$, this implies a proper distance of 121 kpc.  Such a
separation suggests these objects are not interacting, but may be in a
small group.  It is possible that other potential lensed objects in
the system are at the same redshift. Some members of the same group
may lie outside the region magnified by the foreground lensing
cluster.

\section{Discussion}

The large magnification of cB58 makes it an important target for
available instrumentation at many wavelengths.  For example, it is a
planned target for Spitzer guaranteed time observations, which will
probe the star-formation and dust content of such galaxies.  The
addition of the counter arc, A2, and the companion galaxy, W, serve to
increase the return on future observations of the field.  We expect
that confirmation of the C and possibly B sources will soon show them
to be members the same group.

We have shown that A2 is the counter image of MS1512-cB58.  The
spatial resolution provided by differential magnification suggests an
uneven distribution of star-forming regions in the source galaxy.  The
rotational velocity limit obtained from the A2 spectrum supports other
observations of LBGs at somewhat lower redshift.  We have also
confirmed the existence of the companion W source, which is intrinsically fainter than
the A source.  The probable lower metallicity in the W source might be
evidence of the luminosity-abundance relationship at high redshift
(Kobulnicky et al.\ 2003).

These lensed LBGs are valuable examples of moderate luminosity
galaxies.  The magnification enables detailed studies of average
luminosity galaxies at high redshift.  A larger sample is still needed,
however, to draw more general conclusions.  Other galaxy clusters
offer magnified star-forming galaxies at high redshift as well (i.e.\ 
Fosbury et al.\ 2003), but so far none of comparable brightness have
been discovered.  Future study of lensed LBGs will extend to higher
redshifts (i.e.  Ellis et al.\ 2001).  In fact, the observation of
gravitationally lensed sources may reveal the highest redshift and
intrinsically faintest sources (Pell\'{o}\ et al.\ 2003).

\acknowledgements We thank S. Seitz for useful discussions.  The
research described in this paper was carried out, in part, by the Jet
Propulsion Laboratory, California Institute of Technology, and was
sponsored by the National Aeronautics and Space Administration.

\references 

\reference{} Baker, A. J., Lutz, D., Genzel, R., Tacconi, L. J., \& Lehnert, M. D. 2001, A\&A, 372, 37

\reference{} Conselice, C.J., Bershady, M.A., Dickinson, M., \& Papovich, C.\ 2003, AJ, 126, 1183

\reference{} Ellingson, E., Yee, H.K.C,
        Bechtold, J., \& Elston, R. 1996, ApJL 466, 71

\reference{} Ellis, R., Santos, M. R., Kneib, J.-P., Kuijken, K. 2001, ApJ, 560, 119

\reference{} Erb, D. K., Shapley, A. E., Steidel, C. C., Pettini, M., Adelberger, K. L., Hunt, M. P.,
      Moorwood, A. F. M., \& Cuby, J.-G. 2003, ApJ in press, astro-ph/0303392

\reference{} Fosbury, R.A.E., et al.\ 2003, in press, astro-ph/0307162

\reference{} Fruchter, A. S. \& Hook, R. N.  2002, PASP, 114, 144

\reference{} Kennicutt, R. C., Jr.\ 1998, ARA\&A, 36, 189

\reference{} Kobulnicky, H.A., Kennicutt. R.C. Jr., \& Pizagno, J.L. 1999, ApJ, 514, 544

\reference{} Kobulnicky, H. A. \& Zaritsky, D.\ 1999, ApJ, 511, 118

\reference{} Kobulnicky, H. A. \& Koo, D. C.\ 2000, ApJ, 545, 712

\reference{} Kobulnicky, H. A., et al.\ 2003 in press, astro-ph/0305024

\reference{} Lemoine-Busserolle, M., Contini, T., Pell\'{o}, R., Le Borgne, J.-F., Kneib, J.-P., \&
 Lidman, C. 2003, A\&A, 397, 839

\reference{} Lowenthal, J. D., et al.\ 1997, ApJ, 481, 673L

\reference{} Madau, P., Pozzetti, L., \& Dickinson, M. 1998, ApJ, 498, 106

\reference{} McLean, I. S., et al. 1998, SPIE, Vol. 3354, 566

\reference{} McLean, I. S., Graham, J. R., Becklin, E. E., Figer, D. F.,
 Larkin, J. E., Levenson, N. A., \& Teplitz, H. I. 2000, SPIE, 4008, 1048

\reference{} Melbourne, J., \& Salzer, J.J. 2002, AJ, 123, 2302

\reference{} Pagel, B.E.J., Edmunds, M.G., Blackwell, D.E., Chun, M.S. \& Smith, G.
  1979, MNRAS, 189, 95

\reference{} Pell\'{o}, R., et al.\ 2003, to appear in To appear in  "Gravitational Lensing: a unique tool for cosmology",
ASP Conf. S., eds. D. Valls-Gabaud \& J.-P. Kneib; astro-ph/0305229

\reference{} Pettini, M., Kellogg, M., Steidel, C.C., Dickinson, M.,
Adelberger, K.L., \& Giavalisco, M. 1998, ApJ, 508, 539 

\reference{} Pettini, M., Steidel, C.C., Adelberger, K.L., Dickinson,
M., \& Giavalisco, M.  2000, ApJ, 528, 96

\reference{} Pettini, M., Shapley, A. E., Steidel, C. C., Cuby, J.-G., Dickinson, M.,
 Moorwood, A. F. M., Adelberger, K. L., \& Giavalisco, M. 2001, ApJ, 554, 981

\reference{} Pettini, M., Rix, S. A., Steidel, C. C., Adelberger, K. L., Hunt, M. P., \& Shapley, A. E.
 2002, ApJ, 569, 742

\reference{} Seitz, S., Saglia, R.P., Bender, R., Hopp, U., Belloni,
P., \& Ziegler, B.  1998, MNRAS 298, 945 (S98)

\reference{} Steidel, C.C., Giavalisco, M., Pettini, M., Dickinson,
M., \& Adelberger, K.L. 1996, ApJ Letters, 462, L17

\reference{} Steidel, C.C., Adelberger, K.L., Shapley, A.E., Pettini, M., Dickinson, M., Giavalisco, M. 
2003, ApJ in press, astro-ph/0305378

\reference{} Teplitz, H.I. et al.\ 2000a, ApJ, 542, 18

\reference{} Teplitz, H.I. et al.\ 2000b, ApJLetters, 533, 65 (T00)

\reference{} Trager, S.C., Faber, S.M.,Dressler, A., \& Oemler, A.
1997, ApJ, 485, 92

\reference{} Yee, H.K.C., Ellingson, E., Bechtold, R.G.,
Carlberg,R.G., Cuillandre, J.-C. 1996, AJ, 111, 1783

\clearpage

\begin{deluxetable}{lrrrr}

\tablecaption{Photometry \label{tab: phot}}
\tablewidth{0pc}
\tablehead{
\colhead{Object}&
\colhead{F555W$_{AB}$}&
\colhead{F675W$_{AB}$}&
\colhead{F814W$_{AB}$} &
\colhead{K$_{AB}$}

}
\startdata

cB58\tablenotemark{1} & $20.64\pm 0.05$ & $20.41\pm 0.05$ & $20.35\pm 0.05$ & $19.61\pm 0.12$ \\
A2   & $23.26\pm 0.08$ & $23.03\pm 0.08$ & $22.95\pm 0.08$ & $22.09\pm 0.19$ \\                 
W5   & $26.0\pm 0.2$   & $26.1\pm 0.3$   & $26.8\pm 0.3$   & $23.4\pm 0.3$

\enddata

\tablenotetext{1}{Photometric magnitudes are scaled to the values of Ellingson et al.\ (1996), but 
the uncertainties reflect the measurements in this study.  Photometry of A2 and W5 are measured relative
to these data points.}

\end{deluxetable}

\clearpage

\begin{deluxetable}{lrrrrrr}

\tablecaption{Emission Lines \label{tab: emlines}}
\tablewidth{0pc}
\tablehead{
\colhead{Object}&
\multicolumn{3}{c}{[OIII] 5007} &
\multicolumn{3}{c}{\ha} \\
\colhead{} &

\colhead{$\lambda_{obs}(\mu\mbox{m})$}&
\colhead{$W_{rest}\tablenotemark{a}$}&
\colhead{$F\tablenotemark{b}$} &
\colhead{$\lambda_{obs}(\mu\mbox{m})$}&
\colhead{$W_{rest}\tablenotemark{a}$}&
\colhead{$F\tablenotemark{b}$}
}
\startdata

cB58\tablenotemark{c} &  1.86678 & $97\pm 15$   & $14.7\pm 1.5$ & 2.4475 & $106\pm 16$  & $12.6\pm 1.4$  \\
A2                    &  1.86731 & $273\pm 56$ & $4.3\pm 0.6$  & 2.4468 & $261\pm 59$ & $3.1\pm 0.5$ \\
W5                    &  1.86338 & \nodata     & $0.80\pm 0.19$  & 2.4453 & \nodata     & $0.33\pm 0.09$ \\

\enddata
\tablenotetext{a}{Rest-frame equivalent width in \AA}
\tablenotetext{b}{Observed line flux in units of $10^{-16}$\ergcm2s}
\tablenotetext{c}{Data from Teplitz et al.\ (2000)}
\end{deluxetable}

\clearpage

\clearpage

\clearpage

\clearpage

\clearpage

\clearpage

\clearpage

\end{document}